# On the theory of SPASER - a laser with a surface plasmon

V.S. Zuev

The P.N. Lebedev Physical Institute of RAS
119991 Moscow, Russia
vizuev@sci.lebedev.ru

Summary

The paper contains a theory of laser action in a so called SPASER – a laser with a surface plasmon. The goal of the theory development is an ascertainment of physical nature of SPASER lasers that were successfully put into action by M.A.Noginov *et al*, announced in the paper "Demonstration of a spaser-based nanolaser", and by Xiang Zhang *et al*, the paper "Plasmon lasers at deep subwavelength scale", both papers in Nature Advance online publication, August 2009. Plasmonic modes have huge losses but simultaneously the probabilities of radiative processes in atoms (molecules) with emission of radiation into such modes are increased thousandfold as compared with probabilities to radiate in free space. This makes clear how the losses are compensated by amplification. It is made apparent the kinship of physical nature of the phenomenon with that of the SERS.

К теории SPASER'а – лазера с поверхностным плазмоном

В.С.Зуев

Физический институт им. П. Н. Лебедева РАН
119991 Москва, Ленинский пр., 53
zuev@sci.lebedev.ru

Аннотация

В работе предложена теория лазерного действия так называемого SPASER'а – лазера с поверхностным плазмоном наноразмеров. Теория развита с целью выяснения физической природы лазеров типа SPASER'а, о запуске которых сообщено в работах "Demonstration of a spaser-based nanolaser", авторы M.A.Noginov *et al* и "Plasmon lasers at deep subwavelength scale", авторы Xiang Zhang *et al*, обе в Nature, Advance online publication, August 2009. Плазмонные моды наноразмеров имеют чрезвычайно большие потери, но одновременно вероятности радиационных процессов в атомах (молекулах) с излучением в такую моду тысячекратно увеличены по сравнению с этими вероятностями с излучением в свободном пространстве. Указано на родство физической природы явления с природой SERS.

В.С.Зуев

Физический институт им. П. Н. Лебедева РАН
119991 Москва, Ленинский пр., 53
zuev@sci.lebedev.ru

В августе 2009 г. появились две публикации /1,2/, которые заслуживают пристального внимания. В первой из них описан лазер видимого излучения 531 *nm*, представляющий собой металлическую сферу диаметром 14 нм, погруженную в оболочку из кварца с примесью молекул красителя Oregon Green OG-488. Диаметр оболочки 44 нм. Во 2-й работе описан лазер, представляющий собой наноцилиндр из $CdS$ диаметром 100 нм, образующая наноцилиндра параллельна плоской пленке серебра, расстояние до пленки серебра 5 нм, зазор заполнен материалом $MgF_2$. Лазерная генерация возбуждалась вдоль длины наноцилиндра в зазоре между $CdS$ и $Ag$. Площадь поперечного сечения лазерной моды составляла $\lambda^2/400$, $\lambda = 489\ nm$. Длина лазера не указана, но можно показать, что это не существенно. В обоих работах применялось оптическое возбуждение.

Начнем обсуждение с данных работы /1/. Диаметр наносферы, 14 нм, это – половина длины волны плазмона $\lambda_{pl}$. Таким образом волновое число плазмона равно $k_{pl} = \dfrac{2\pi}{\lambda_{pl}} = 2.244 \cdot 10^6\ cm^{-1}$. Это – действительная часть. Мнимую часть определим из приведенной в статье цифры для добротности резонанса $Q$. На длине волны $\lambda = 525\ nm$ эта добротность указана равной $Q = 14.8$. Добротность плазмонного резонанса равна $Q = \dfrac{\operatorname{Re} k_{pl}}{\operatorname{Im} k_{pl}}$, откуда для $\operatorname{Im} k_{pl} = \operatorname{Re} k_{pl}/Q$ получаем значение $1.516 \cdot 10^5\ cm^{-1}$.

Полученное значение $\operatorname{Im} k_{pl} = 1.516 \cdot 10^5\ cm^{-1}$, значение показателя поглощения для распространения поверхностного плазмона, чрезвычайно велико. Показатели усиления в растворах красителей и в полупроводниках едва ли превышают значение 100 $cm^{-1}$.

Однако, следует принять во внимание, что приведенное значение показателя усиления относится к однородной среде и к однородному пространству. В пространстве с наносферой вероятности радиационных переходов и пропорциональные им показатели усиления сильно увеличены. Применительно к комбинационному рассеянию эффект получил название SERS – Surface Enhanced Raman Scattering. Применительно к однофотонному радиационному переходу эффект имеет ту же природу. Он рассмотрен нами в работе /3/.

Физическая природа эффекта заключается в том, что вероятность радиационных переходов не является величиной универсальной, а зависит от структуры пространства. Вблизи металлической наночастицы вероятность радиационного перехода увеличена в сотни и тысячи раз /3/.

Будем следовать работе /3/. Рассмотрим плазмоны на наноцилиндре малой длины. Это нагляднее, чем рассмотрение сферы. Из этого рассмотрения возникают приближенные выражения для наносферы и наноэллипсоида.

Среди многочисленных поверхностных мод цилиндра только мода, содержащая $\vec{n}_{0\gamma} e^{ihz}$, так называемая $TM_0$ мода не имеет критической частоты: она существует на цилиндрах сколь угодно малого диаметра. Именно эта мода определяет рассматриваемый эффект лазерной генерации. Будем считать радиус цилиндра равным $a$, все физические величины и переменные для пространства вне цилиндра будем отмечать индексом $i = 1$, внутри цилиндра – индексом $i = 2$. Поле этой моды имеет вид /4/

$$\vec{N}_{0\gamma} = \vec{n}_{0\gamma} e^{ihz},\ \vec{n}_{0\gamma} = \frac{ih}{\sqrt{k^2}} \frac{d}{dr} Z_0(\gamma r) \cdot \vec{i}_r - \frac{\gamma^2}{\sqrt{k^2}} Z_0(\gamma r) \cdot \vec{i}_z. \tag{15}$$

Здесь $\gamma = \sqrt{h^2 - k^2}$, $k^2 = \varepsilon \mu (\omega/c)^2$, $Z_0(\gamma r) = I_0(\gamma r)$, $K_0(\gamma r)$ - модифицированные бесселевы функции. Выбирая векторный потенциал в виде $\vec{A}_i = A_i \vec{N}_{0\gamma_i} e^{-i\omega t}$, вычисляя поле $\vec{E}$ и $\vec{H}$ и затем приравнивая при $r = a$ тангенциальные компоненты $\vec{E}$ и $\vec{H}$ вне и внутри цилиндра получаем следующее характеристическое уравнение:

$$\frac{\gamma_2}{\gamma_1}\frac{I_0(\gamma_2 a)}{K_0(\gamma_1 a)} = -\frac{\varepsilon_2}{\varepsilon_1}\frac{I_1(\gamma_2 a)}{K_1(\gamma_1 a)}. \qquad (16)$$

Когда $a$ мало, тогда $TM_0$ - волна оказывается сильно замедленной в сравнении с плоской волной в свободном пространстве.

Теперь мы имеем все необходимое для вычисления вероятности излучательного перехода с испусканием фотона в поверхностную волну. Мы будем следовать представлениям квантовой теории излучения, изложенной в /5/. Вероятность спонтанного излучательного перехода описывается формулой

$$w_{n|0} = \frac{2\pi}{\hbar}|H_{n|0}|^2 \rho_{E_n}. \qquad (24)$$

Вероятность $w_{n|0}$ имеет размерность, обратную времени, матричный элемент оператора взаимодействия $H_{n|0}$ - размерность энергии, плотность состояний поля $\rho_E$ - размерность, обратную энергии. Задача состоит в вычислении матричного элемента $H_{n|0}$ и в определении плотности состояний $\rho_E$. Процедура состоит в представлении поля в виде собственных волн - радиационных осцилляторов поля рассматриваемого пространства, которые квантуют. Рассматривая однородное пространство обычно представляют поле в виде набора плоских волн. Можно применить разложение по цилиндрическим либо по сферическим волнам. Результат вычислений не зависит от выбора представления поля, на что обращено специальное внимание в /5/, хотя и без конкретной демонстрации этого свойства. Для неоднородного пространства рассмотрение требуется проделать вновь.

Результат вычислений, подробности которых изложены в /3/, состоит в следующем. Для атомарного диполя $\vec{p}$, находящегося в точке и ориентированного вдоль $\vec{i}_r$ вероятность спонтанного излучательного перехода в $TM_0$ - волну оказывается равной

$$w^{(TM_0)}_{an'_\lambda=1|bn_\lambda=0} = F_1 \cdot F_2, \quad F_1 = \frac{4}{3}\frac{e^2\omega^3}{\hbar c^3}|x_{ab}|^2, \quad F_2 = \frac{3h_0}{C_1+C_2}R_A^2\left(\frac{h_0\gamma_{01}}{k_{01}}\right)^2 K_1^2(\gamma_{01}a_0)\frac{\operatorname{Re}h}{\operatorname{Im}h}. \qquad (32)$$

Константы $C_1$ и $C_2$ возникают при нормировании потенциала, константа $R_A$ возникает из граничных условий:

$$\int_V [\vec{A}(\vec{r})]^2 d\tau = \int_{\substack{Outer\,Space\\i=1}}[\vec{A}_1(\vec{r})]^2 d\tau + \int_{\substack{Metal\\i=2}}[\vec{A}_2(\vec{r})]^2 d\tau = \frac{4\pi^2|A_2|^2}{k_0 h_0}(C_1+C_2) = 4\pi c^2,$$

$$C_1 = R_A^2\left[(h_0/k_{01})^2\int_{\gamma_{01}a_0}^\infty x[K_1(x)]^2 dx + (\gamma_{01}/k_{01})^2\int_{\gamma_{01}a_0}^\infty x[K_0(x)]^2 dx\right],$$

$$C_2 = (h_0/\kappa_{02})^2\int_0^{\gamma_{02}a_0} x[I_1(x)]^2 dx + (\gamma_{02}/\kappa_{02})^2\int_0^{\gamma_{02}a_0} x[I_0(x)]^2 dx,$$

$$R_A = \frac{|A_1|}{|A_2|} = \left(\frac{\gamma_2}{\gamma_1}\right)^2\frac{k_1}{\kappa_2}\frac{I_0(\gamma_2 a)}{K_0(\gamma_1 a)}.$$

$$k_0 = \omega/c, \ k_1^2 = \varepsilon_1\mu_1(\omega/c)^2, \ k_2^2 = \varepsilon_2\mu_2(\omega/c)^2, \ \kappa_2 = \sqrt{-k_2^2},$$
$$h_0 = h/k_0, \ k_{01} = k_1/k_0, \ k_{02} = k_2/k_0, \ \kappa_{02}^2 = -k_{02}^2, \ \gamma_{01} = \gamma_1/k_0, \ \gamma_{02} = \gamma_2/k_0.$$

Фактор $F_1$ совпадает с выражением для вероятности спонтанного излучательного перехода в свободном пространстве. Фактор $F_2$ показывает, какие изменения происходят в пространстве с наноцилиндром. Подстановка значений параметров для опыта из /1/ дает следующее значение $F_2$:

$$F_2 = 2.675 \cdot 10^3 \frac{\operatorname{Re} h}{\operatorname{Im} h}$$

для $2a = 14 \ nm$, $\varepsilon_1 = 2.25$, $\varepsilon_2 = -4.5$, $\lambda = 525 \ nm$. Множитель $\frac{\operatorname{Re} h}{\operatorname{Im} h}$ есть добротность моды. В /1/ эта величина выбрана равной 14.8. Наши вычисления, основанные на данных из /6/, дают $\frac{\operatorname{Re} h}{\operatorname{Im} h} = $. Авторы /1/ также используют данные из /6/. В чем причина разницы – не понятно. Впрочем, эта разница не существенна. Будучи умноженным на 100 $cm^{-1}$ этот показатель усиления перекрывает показатель потерь в плазмоне. Лазерная генерация возможна, что и продемонстрировано в эксперименте /1/.

Поверхностный плазмон на наноцилиндре бесконечной длины не обнаруживает себя во внешнем пространстве. Его поле убывает экспоненциально при удалении от наноцилиндра. А полуволновый отрезок наноцилиндра с плазмоном излучает как элементарный диполь /3/. Дипольная мода оказывается связанной с модой поверхностного плазмона. Для малой наночастицы $2a \ll \lambda$ излучательные потери малы в сравнении с материальными потерями. С ростом размера наночастицы они возрастают.

Постановка опыта в работе /2/ внешне сильно отличается от той, что представлена в работе /1/. Но несмотря на внешнее различие физика процессов в обоих случаях одинакова. Во 2-м случае нет нанотела, но есть электромагнитная мода наноразмеров. Для излучающего атома важно наличие не тела, а моды наноразмера.

Мода образуется следующим образом. Заметим, что малые, толщиной 1-5 нм прослойки из $MgF_2$, как в данной работе, и из sodium silicate в /1/, можно не принимать во внимание. По-видимому, они нужны из соображений химической стойкости. При рассмотрении электромагнитных полей их можно считать отсутствующими. Цилиндр из $CdS$ и пленка $Ag$ образуют пространство с одиночной плоской границей раздела. Вблизи поверхности можно не учитывать кривизну цилиндра из $CdS$. Наличие этой кривизны приводит лишь к боковому ограничению плазмонной волны, движущейся вдоль линии касания (или почти касания) цилиндра и плоскости. На плоской границе имеется поверхностный плазмон с волновым числом $k_{pl} = \frac{\omega}{c}\sqrt{\frac{\varepsilon_1\varepsilon_2}{\varepsilon_1+\varepsilon_2}}$. При $\varepsilon_1 > 1$, $\varepsilon_2 < 0$, $|\varepsilon_2| > \varepsilon_1$ и $|\varepsilon_2| \approx \varepsilon_1$ оказывается, что $k_{pl} \gg \omega/c$. Имеется очень медленный поверхностный плазмон. Указанные равенства и неравенства, похоже имеют место для пары $CdS/Ag$ на длине волны 489 нм. Боковое, в плоскости границы раздела сред ограничение плазмона вполне естественно, так как существуют гауссовы пучки поверхностных плазмонов. Возникает пространственная мода в виде вытянутого эллипсоида. На этой моде и происходит генерация лазерного излучения. Во вне излучение выходит через торцы области, которые не являются идеально отражающими.

Заканчивая рассмотрение работы /2/ отметим два обстоятельства, вызывающие возражение. Во 1-ых, нет нужды называть возбуждаемую моду гибридной. Никакой гибридизации с модой диэлектрического волокна не происходит. При диаметре волокна 100 нм волноводные моды в волокне отсутствуют /2/. Во 2-ых, отсутствие наблюдаемого порога генерации не означает реальное его отсутствие. Нет никаких оснований предполагать этот порог отсутствующим.

В заключение отметим следующее. В так называемых SPASER'ах возбуждаются плазмонные моды, размер которых может быть как угодно мал, пока вещество можно считать непрерывной средой. Потери в плазмонных модах чрезвычайно велики – до $10^4$ $cm^{-1}$ и более. Однако вероятность радиационного процесса в атоме (молекуле, квантовой точке) с излучением в плазмонную моду наноразмера сто- и тысячекратно увеличена в сравнении с вероятностью такого же процесса в однородном пространстве без поверхностных волн. Такое же возрастание испытывает действующее значение коэффициента усиления возбужденной среды. В результате оказывается возможной компенсация потерь и лазерная генерация на поверхностной плазмонной моде.

Аннотация

В работе предложена теория лазерного действия так называемого SPASER'а – лазера с поверхностным плазмоном наноразмеров. Теория развита с целью выяснения физической природы лазеров типа SPASER'а, о запуске которых сообщено в работах "Demonstration of a spaser-based nanolaser", авторы M.A.Noginov *et al* и "Plasmon lasers at deep subwavelength scale", авторы Xiang Zhang *et al*, обе в Nature, Advance online publication, August 2009. Плазмонные моды наноразмеров имеют чрезвычайно большие потери, но одновременно вероятности радиационных процессов в атомах (молекулах) с излучением в такую моду тысячекратно увеличены по сравнению с этими вероятностями с излучением в свободном пространстве. Указано на родство физической природы явления с природой SERS.

Summary

The paper contains a theory of laser action in a so called SPASER – a laser with a surface plasmon. The goal of the theory development is an ascertainment of physical nature of SPASER lasers that were successfully put into action by M.A.Noginov *et al*, announced in the paper "Demonstration of a spaser-based nanolaser", and by Xiang Zhang *et al*, the paper "Plasmon lasers at deep subwavelength scale", both papers in Nature Advance online publication, August 2009. Plasmonic modes have huge losses but simultaneously the probabilities of radiative processes in atoms (molecules) with emission of radiation into such modes are increased thousandfold as compared with probabilities to radiate in free space. It is made apparent the kinship of physical nature of the phenomenon with that of the SERS.